\documentclass[aps,superscriptaddress,amsfonts,amssymb,amsmath,showpacs,nofootinbib, twocolumn, floatfix]{revtex4-1}
\usepackage{graphicx}
\usepackage[squaren]{SIunits} 
\usepackage[usenames]{color}
\usepackage{url} 
\newcommand{\Eref}[1]{Eq.~\eqref{#1}} 
\newcommand{\Eqnsref}[2]{Equations~\eqref{#1} and~\eqref{#2}} 
\newcommand{\Fref}[1]{Fig.~\ref{#1}} 
\newcommand{\Sref}[1]{Sec.~\ref{#1}}

\begin{document}

\title{R-mode frequencies of rapidly and differentially rotating relativistic neutron stars}

\author{Michael Jasiulek} 
\email{e-mail: michael.jasiulek@ufabc.edu.br} 

\author{Cecilia Chirenti} 
\email{e-mail: cecilia.chirenti@ufabc.edu.br} 
\affiliation{Centro de Matem\'atica, Computa\c c\~ao e Cogni\c c\~ao, UFABC, 09210-170 Santo Andr\'e-SP, Brazil}

\begin{abstract}
R-modes of neutron stars could be a source of gravitational waves for ground based detectors.
If the precise frequency $\sigma$ is known, guided gravitational wave searches with enhanced detectability
are possible. Because of its physical importance many authors have calculated the
r-mode frequency. For the dominant mode, the associated gravitational wave frequency is 4/3 times the angular velocity 
of the star $\Omega$, subject to various corrections of which relativistic and rotational corrections
are the most important.
This has led several authors to investigate the dependence of the r-mode frequency on 
factors such as the relativistic compactness parameter ($M/R$) and the angular velocity of stars with different equations of state. The results found so far, however, are almost independent of the equation of state. 
Here we investigate the effect of rapid rotation and
differential rotation on $\sigma$. We evolve the perturbation equations using the Cowling
approximation by applying finite differencing methods to compute the r-mode frequency
for a series of rotating neutron stars with polytropic equations of state. 
We find that rotational effects in the r-mode frequency can be larger than relativistic effects for rapidly spinning  stars with low compactness, reducing the observed frequency. Differential rotation also acts to  decrease $\sigma$, but its effect inscreases with the compactness. The results presented here are relevant to the design of gravitational wave and
electromagnetic r-mode searches. 
\end{abstract}


\maketitle

\section{Introduction}
\label{sec:intro}

The gravitational-wave driven Chandrasekhar-Friedman-Schutz (CFS) 
instability \cite{Chandrasekhar,Friedman} associated with r-modes \cite{Papaloizou} in
rotating neutron stars has attracted considerable attention in 
recent years. The r-mode gravitational waves could provide an explanation
for the low spin-rates of young neutron stars as they provide a 
spindown mechanism that could explain why
accreting pulsars in Low Mass X-ray Binaries (LMXB) do not spin up to their braking-frequency.
Moreover, r-mode gravitational waves may be detectable by groundbased
detectors in the near future \cite{Owen}, although there is still much discussion about the r-mode saturation amplitude, see for instance \cite{Arras, Bondarescu1, Bondarescu2, Haskell}.
Despite the possibly low saturation amplitude, r-modes could even be observed through their
electromagnetic signatures in burst oscillations of LMXB, see \cite{Strohmayer1,Strohmayer2}. 

Nearly 30 years of observations of binary pulsars and LMXB reveal 
rotation frequencies up to 716Hz \cite{Manchester}. For these pulsars, rotational effects on the r-mode frequency 
have to be considered. 

It is also believed that accretion as well as r-mode oscillations could 
drive a star into differential rotation, while magnetic braking
can damp this effect \cite{Rezzolla1,Rezzolla2}. Still, a small amount of 
differential rotation may be present in all LMXB pulsars and its effect on the r-mode
frequency could be relevant. Differential rotation can also  be an important ingredient when considering very young neutron stars or hypermassive neutron star remnants from binary mergers (see \cite{Kastaun1} for a recent study).

The effect of differential rotation on the frequencies of stellar modes of oscillation in the Cowling approximation was already investigated in a more preliminary way in \cite{Stavridis, Passamonti1} (f and p modes) and \cite{Chirenti} (f and r-modes). (Oscillations of fast and differentially rotating stars in the Cowling approximation were considered in \cite{Krueger}, but their analysis was mostly focused on the f-mode instability.) We present here a first study that incorporates fast rotation and a (possibly) strong degree of differential rotation in the analysis of the r-mode frequencies.

We note here that only a few perturbative studies of rotating neutron stars have been performed without employing the Cowling approximation. Generic inertial modes were considered in \cite{Lockitch}, r-modes in slowly and uniformly rotating stars were studied in \cite{Idrisy}, and in \cite{Ferrari} a general perturbative formalism using spectral methods was presented and applied to f-modes. We also present here a first quantitative assessment of the accuracy of the Cowling approximation for r-modes, comparing our results with those from \cite{Idrisy}.

The structure of the paper is as follows. In Section \ref{sec:corrections} we present a parametrization for the relativistic and rotational corrections to the r-mode frequency, including corrections from the differential rotation. Our numerical results are detailed and analyzed in Section \ref{sec:results}, and we close with our final remarks in Section \ref{sec:conclusion}. Unless otherwise explicitely stated, we use units in which $c = G = M_{\bigodot} = 1$.

\section{Corrections to the r-mode frequency}
\label{sec:corrections}

The observed gravitational wave r-mode angular frequency $\sigma$ is related
to the rotating frame r-mode frequency $\sigma_R$ by  
\begin{equation}
   \sigma = (\kappa - 2) \Omega = \sigma_R - 2\Omega,\quad \kappa \equiv \frac{\sigma_R}{\Omega}, 
   \label{eq:sigma}
\end{equation}
here for the dominant $l=m=2$ r-mode, where $\Omega=2\pi f$ is the angular velocity 
of the star and $\kappa$ is the dimensionless r-mode frequency in the rotating frame \cite{Stergioulas}. 
For a slowly rotating Newtonian star $\kappa=2/3$, independent of the equation of state (EoS) \cite{Papaloizou}. 

Relativistic effects are quantified by the dimensionless compactness $M/R$
of the star. For slowly rotating polytropic stars over a range of astro-physically
motivated compactnesses $0.11-0.21$ the authors of \cite{Idrisy} found 
\begin{eqnarray}
   \kappa_{\textrm{slow rotation}} &\equiv& \kappa_0 = 0.616 + \nonumber \\
&+& 0.352 (M/R) - 3.47 (M/R)^2,
   \label{eq:kappaxM/R}
\end{eqnarray}
corresponding to a correction from its Newtonian value of $8\%-20\%$. Note that
$\kappa_0$ is decreased and thus $\sigma$ is increased with increasing compactness. 

The effect of fast stellar rotation including the leading order correction to the r-mode frequency was considered by \cite{Lindblom1, Passamonti2} 
and can be expressed as
\begin{equation}
  \kappa = \kappa_0 + \kappa_2\, \frac{\Omega^2}{\pi G \bar{\rho}_0}\,,
  \label{eq:kappaxrot}
\end{equation}
where $\bar{\rho}_0$ is the average density obtained numerically as (total mass)/(total volume), and  $\kappa_2$ is
dimensionless and of order unity. 

Since for rotating stars there is no unique radius to compute the
compactness, we take here $R$  to be the volumetric radius, i.e., the radius of the sphere that has the same volume as the star. For a typical compactness of $0.15$ the correction is $9\%-17\%$ for stars with typical radius
from $R=10\,\textrm{km}$ to $R=14\,\textrm{km}$. 

In order to consider the influence of differential rotation in the r-mode frequencies, we use the so-called {\it j-const. law} derived by \cite{Komatsu1,Komatsu2}, that has been widely used in the literature. (For more general rotation laws, see \cite{Galeazzi,Mach}.) In the Newtonian limit, the j-const. law gives the angular velocity $\Omega$ of the star as
\begin{equation}
 \Omega = \frac{\Omega_c}{1 + \hat{A}^{-2}\hat{r}^2 \sin^2\theta}\,,
 \label{eq:j-const}
\end{equation}
where $\Omega_c$ is the central angular velocity and $\hat{A} \equiv A/R_e$ and $\hat{r} \equiv r/R_e$ are dimensionless quantities rescaled in terms of the equatorial radius $R_e$, following the notation in \cite{Baumgarte}. The parameter $A$ controls the degree of differential rotation, but we will follow the literature and use the dimensionless parameter $\hat{A}$, such that stars with different radii but the same $\hat{A}$ have the same amount of differential rotation.

In \Sref{sec:results} below we analyze the variation of $\kappa_2$ with the compactness $M/R$ and we generalize \Eref{eq:kappaxrot} to include the differential rotation correction as an additional term, $\kappa_3$. As we increase the number of degrees of freedom in our stellar model (compactness, total angular momentum and increasing differential rotation), the simplicity of the Newtonian formula for the r-mode frequency is somewhat lost. However, we gain finer and more detailed information on the frequencies, with non-negligible corrections that could change the value of $\kappa$ by almost a factor 2.

\section{Numerical results}
\label{sec:results}
\subsection{Fast rotation}
 
We evolve the perturbation equations using a finite-differencing code 
in $3+1$ dimensions within the \texttt{Einstein Toolkit} \cite{ET}. We use an
outer 6-patch spherical coordinate system with an inner Cartesian cube
to avoid coordinate singularities. This coordinate setup is provided by the
extension Llama, see \cite{Pollney} or Sec. 5.1 from \cite{Jasiulek} for code tests 
and more details.
The background models in the finite-differencing code are represented
using spectral methods and computed with the \texttt{Lorene Code} \cite{Lorene}. We use
the same scaling function as Lorene to adapt the surface to
a (round) sphere for rotating models. 
Both codes are available online and we also plan to make our code freely
available. 

\begin{figure}[!htb]
\begin{center}
\includegraphics[width=0.98\columnwidth]{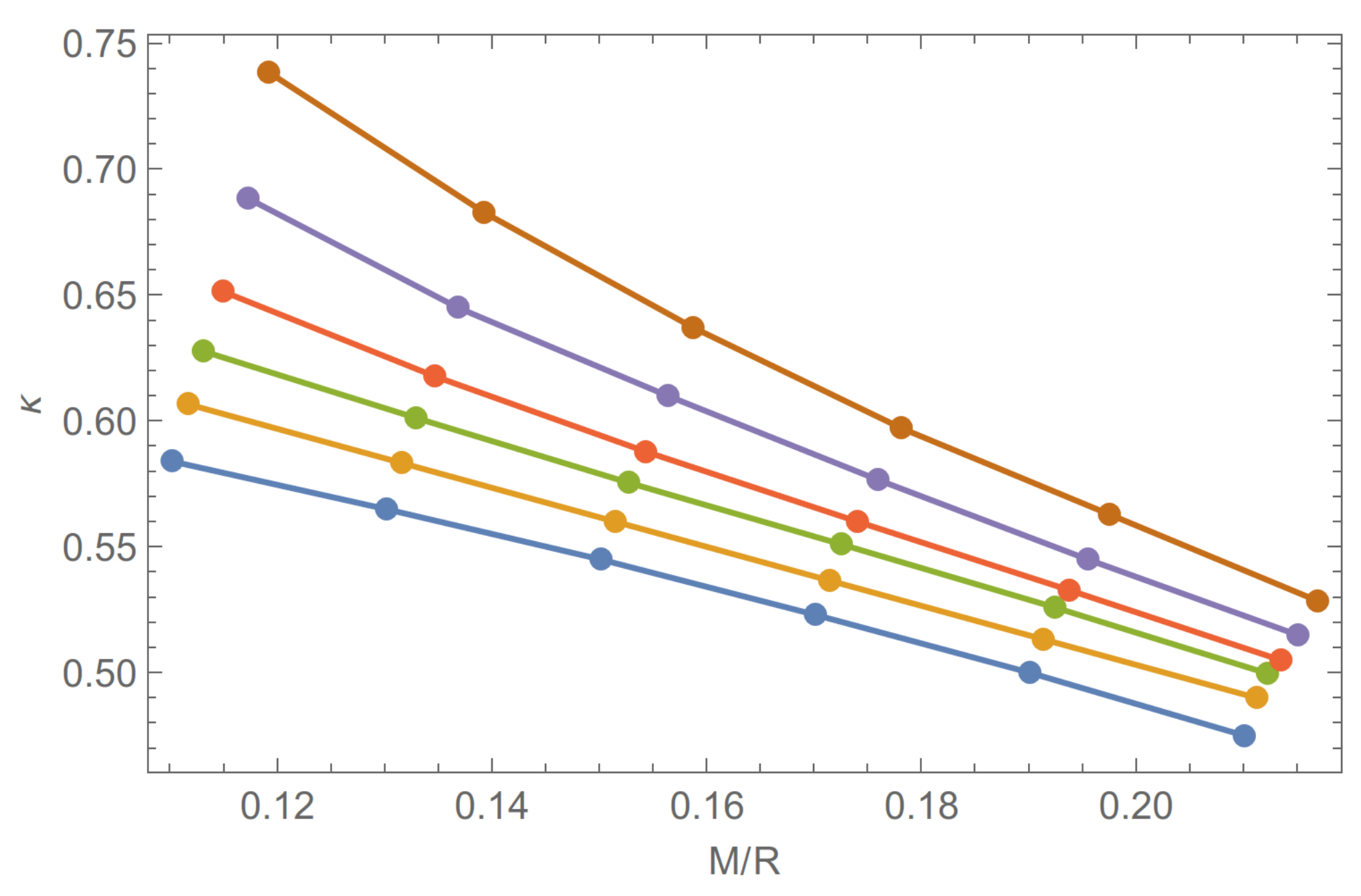}
\end{center}
\caption{$\kappa \times M/R$ for sequences of stars with different values of $f$. From top to bottom, $f = 700, 600, 500, 400, 300,100$ Hz. Note that $\kappa$ decreases (and $\sigma$ increases) for more compact stars, but $\kappa$ increases (and $\sigma$ decreases) for more rapidly rotating stars.}
\label{fig1}
\end{figure}

We ran our code for several sequences of polytropic stars with fixed polytropic index $N=1$ and total radius $R = 14.15$,  compactness varying from $0.11-0.21$, and uniform rotation
in the range between $100\,\textrm{Hz} - 700\,\textrm{Hz}$. 
In our first code tests we compared results with values in \cite{Chirenti, Ruoff, Kastaun2}, with very good agreement. 
We obtained $\kappa$ as a function of both $M/R$ and $\Omega$, see \Fref{fig1}. In agreement with 
\Eqnsref{eq:kappaxM/R}{eq:kappaxrot}, we can see that $\kappa$ is increasing with increasing rotation and decreasing with increasing
compactness of the star. Within our chosen range of parameters, $\kappa$ increases with rotation by at least 30\% for low compactness ($M/R = 0.12$) and by less than 13\% for high compactness ($M/R = 0.22$).

\begin{figure}[!htb]
\begin{center}
\includegraphics[width=0.98\columnwidth]{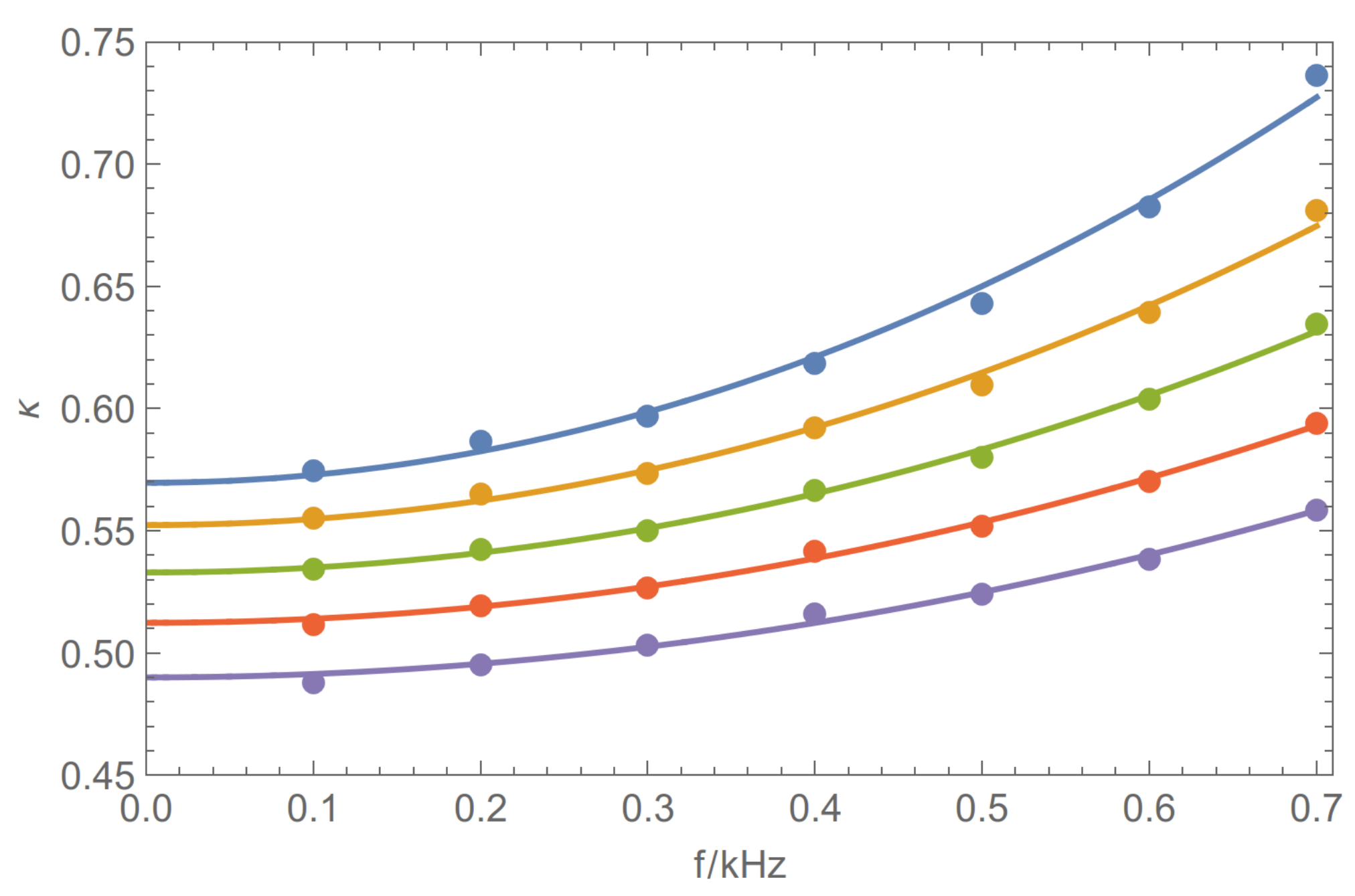}
\end{center}
\caption{$\kappa \times f$ for sequences of stars with different values of $M/R$, which were obtained by interpolation from the data in \Fref{fig1} (dashed lines). From top to bottom, $M/R = 0.12, 0.14,0.16,0.18,0.20$. The solid lines represent quadratic fits of the form $\kappa = a + bf^2$, where both $a$ and $b$ decrease with the compactness.}
\label{fig2}
\end{figure}

In \Fref{fig2} we fitted our data to $\kappa=a+b\,f^2$, where $a \equiv \kappa_0$ (for a given compactness) and $b \equiv \kappa_2/(\pi G\bar{\rho}_0)$, see \Eref{eq:kappaxrot}. We can see that fit and data show very good agreement and that $a$ and $b$ (i.e. $\kappa_0$ and $\kappa_2$) are both decreasing with increasing compactness. This trend, which was already known for $\kappa_0$, is to be expected because both the gravitational redshift and the frame dragging will act to decrease the oscillation frequencies as measured by a comoving observer (see eq.(30) in \cite{Lockitch}).

\begin{figure}[!htb]
\begin{center}
\includegraphics[width=0.98\columnwidth]{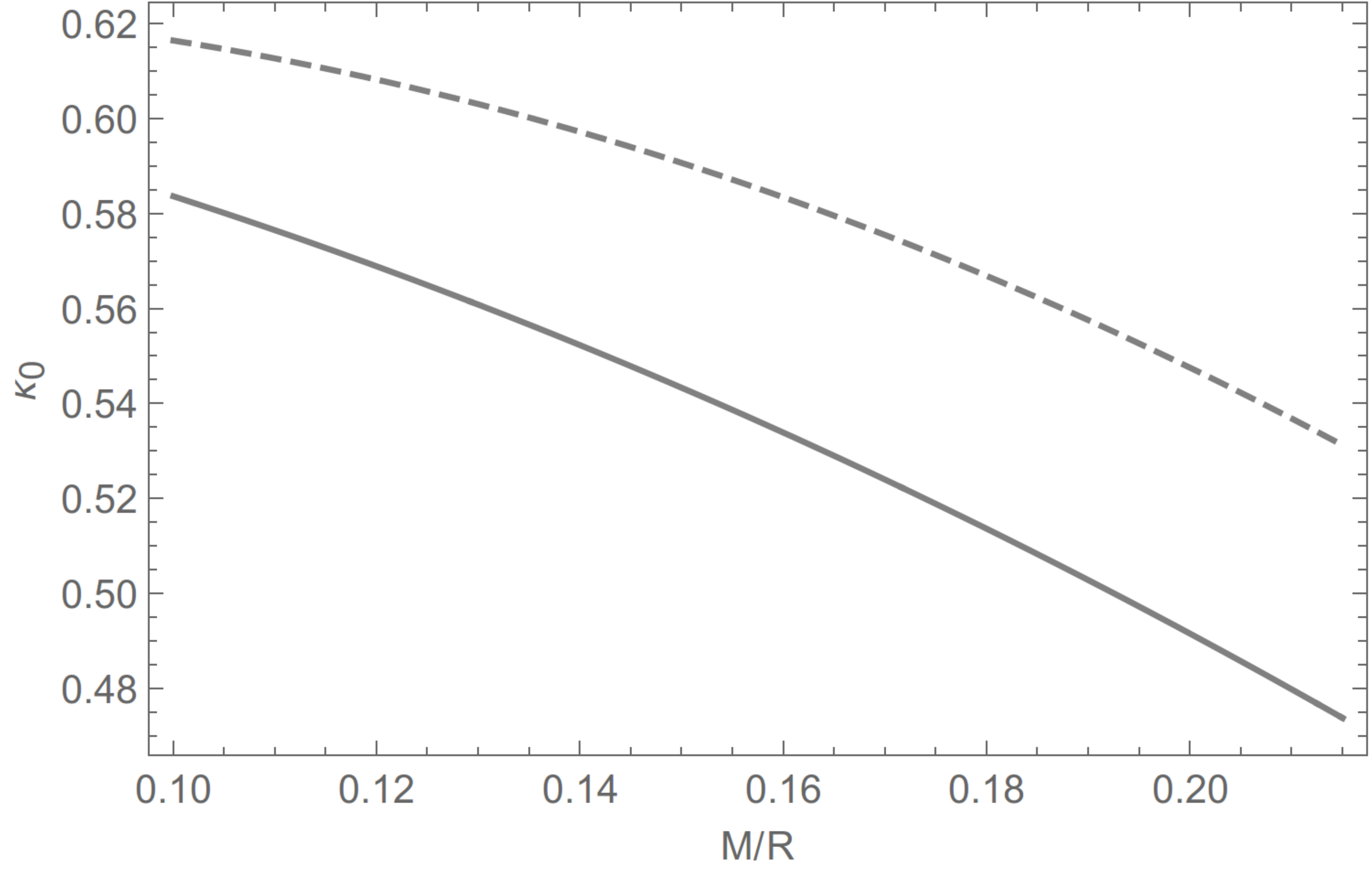}
\end{center}
\caption{$\kappa_0 \times M/R$. The dashed line represents \Eref{eq:kappaxM/R} and the solid line is given by a quadratic fit of the parameter $a$ in \Fref{fig2} that gives $\kappa_0 = 0.64 - 0.35(M/R) - 2.0(M/R)^2$. The difference between the two curves shows the accuracy of the Cowling approximation.}
\label{fig3}
\end{figure}

The quadratic fit for $\kappa_0 = \kappa_0(M/R)$ should be compared with \Eref{eq:kappaxM/R} and both curves are shown in \Fref{fig3}. The difference between the two curves in \Fref{fig3} shows the accuracy of the Cowling approximation for r-modes, increasing with compactness and ranging from $6\%-11\%$ for the models we have considered. To the best of our knowledge, a quantitative measurement of this error has not been presented in the literature so far, and confirms the usual intuition that the Cowling approximation should provide reasonably accurate results for r-modes (for a study on the accuracy of the Cowling approximation for f-modes, see \cite{Yoshida}).

\begin{figure}[!htb]
\begin{center}
\includegraphics[width=0.98\columnwidth]{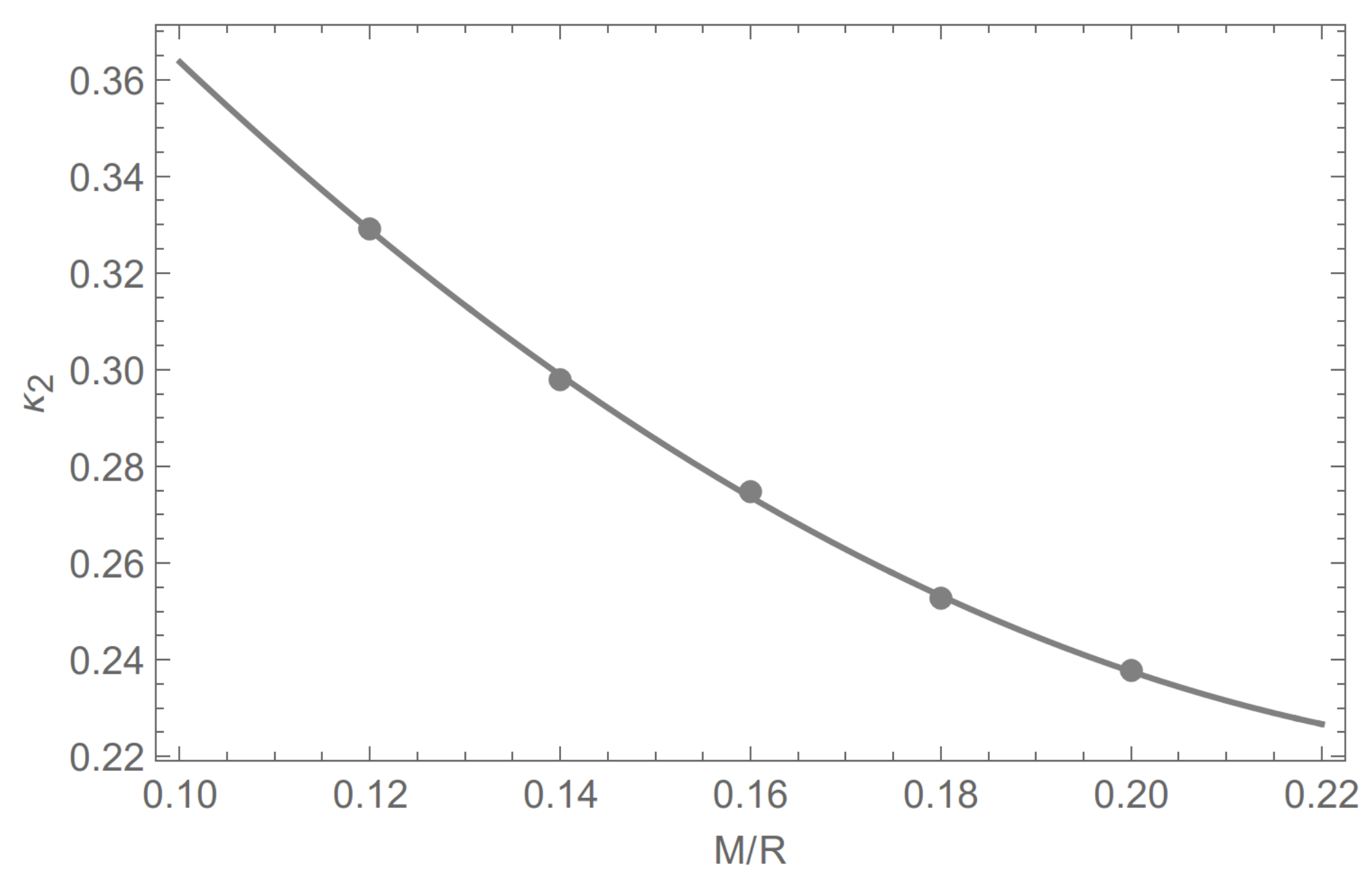}
\end{center}
\caption{$\kappa_2 \times M/R$. The solid lines shows a quadratic fit of $\kappa_2$ obtained from the parameter $b$ in \Fref{fig2} that gives $\kappa_2 = 0.68 - 4.1(M/R) + 9.8(M/R)^2$. The rotational correction $\kappa_2$ is more significant for less compact stars.}
\label{fig4}
\end{figure}

A similar quadratic fit for $\kappa_2 = \kappa_2(M/R)$ can be seen in \Fref{fig4}. Our results show that the dependence of $\kappa_2$ with the compactness should not be neglected, since the difference between values for low and high compactness is almost 70\%. We also note here that our values for $\kappa_2$ agree well with the representative value 0.29 given in \cite{Idrisy,Lindblom1}.

\subsection{Differential rotation}

\begin{figure}[!htb]
\begin{center}
\includegraphics[width=0.98\columnwidth]{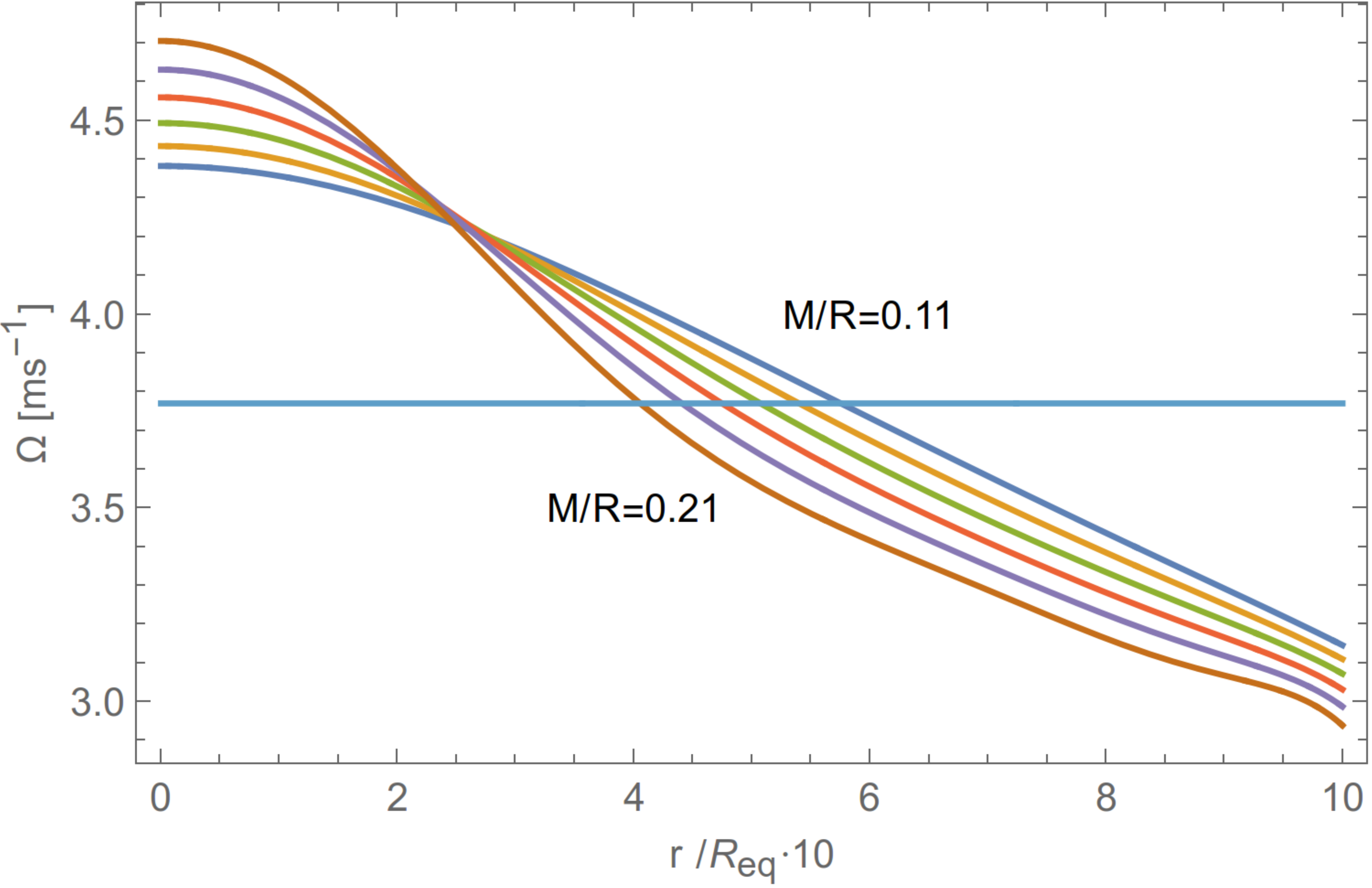}
\end{center}
\caption{The rotation profiles as a function of the radial coordinate scaled by the equatorial radius $R_{\rm eq}$ of 6 differentially rotating stars with $\hat{A}^{-1} = 0.5$ that have the same total angular momentum as the corresponding uniformly rotating models that have the same compactness and
$f=600Hz$. We see that although the differential rotation parameter across the models is set 
constant, stars with higher compactness have slightly more physical differential rotation. $M/R = 0.11, 0.13, ...\,, 0.21$ as indicated in the figure.}
\label{6-models}
\end{figure}

We now allow our stars to have differential rotation, as described by the j-const. law. As we have already discussed in \Sref{sec:corrections}, the degree of differential rotation increases with increasing $\hat{A}^{-1}$. However, the situation is not so simple, as can be seen in \Fref{6-models} for a representative case: for a fixed differential rotation parameter, more compact stars will present a larger variation between the central and the surface angular velocity in the equatorial plane. This variation in $\Omega$ can be understood as a more natural measure of the ``physical'' differential rotation of the star, and it ranges from 30\% to 38\% for the models presented in \Fref{6-models}. 

All models shown in \Fref{6-models} have the same total angular momentum as the uniformally rotating model with $f = 600\,\textrm{Hz}$. Moreover, $\Omega(r,\theta)$ of the differentially rotating models have volume averages about the angular velocity $\Omega$ of the uniformly rotating model (represented by a horizontal line), as can be seen in \Fref{6-models}.\footnote{The small ``bump" in $\Omega$ close to the surface of the most compact star (the star with the strongest differential rotation) in \Fref{6-models} is a feature of the scalling function we use for adapting the surface of our stars to a round sphere \cite{Lorene}. } Therefore we will use this $\Omega$ to define $\kappa$ in the differentially rotating case, in analogy with \Eref{eq:sigma}. 

\begin{figure}[!htb]
\begin{center}
\includegraphics[width=0.98\columnwidth]{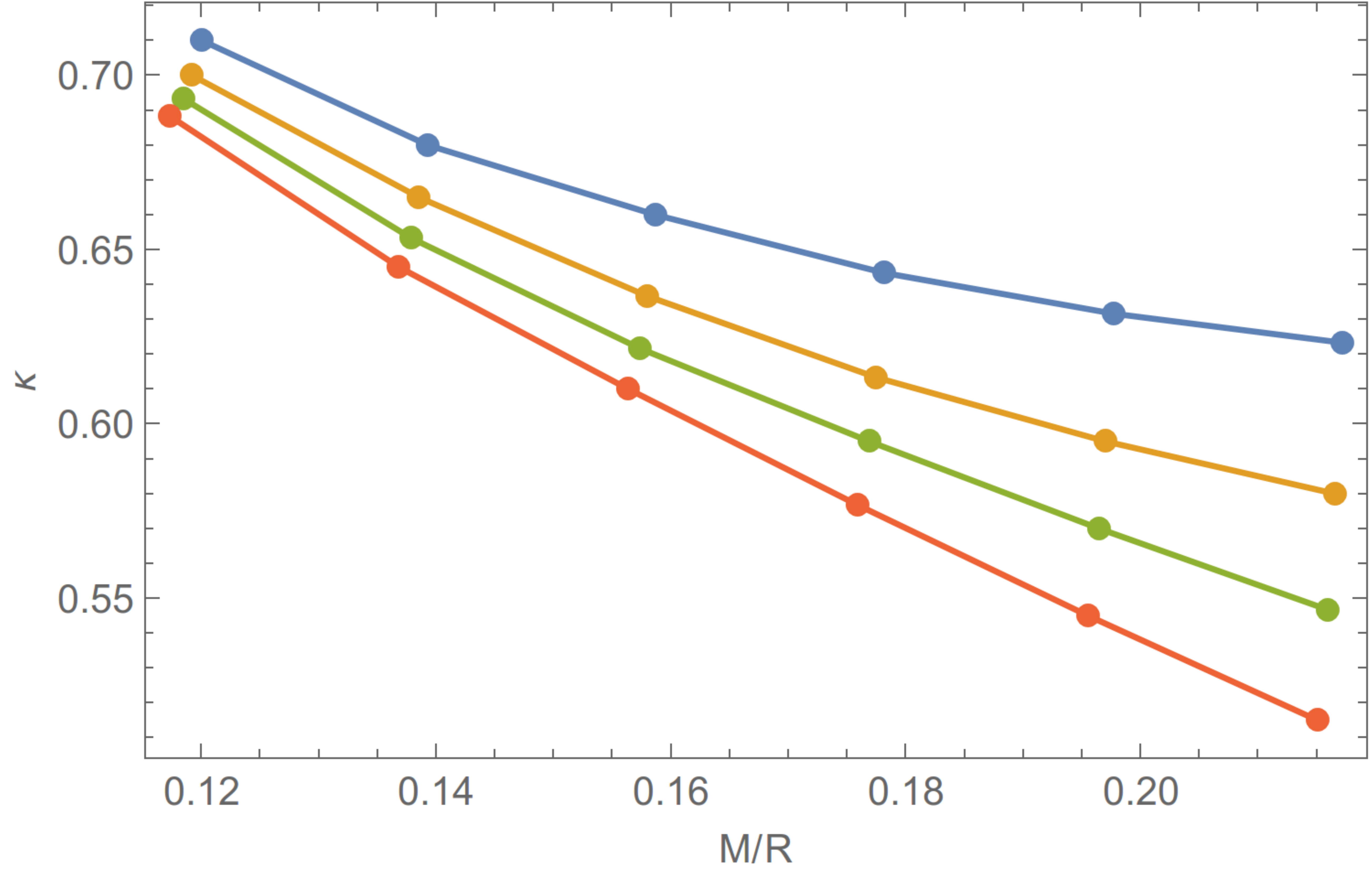}
\end{center}
\caption{$\kappa \times M/R$ for sequences of stars with different values of the differential rotation parameter $\hat{A}^{-1}$. The upper curve shows $\kappa$ for the models presented in \Fref{6-models}. The other curves are similar, but with $\hat{A}^{-1} = 0.4,0.3$ and 0.0 (from top to bottom). Note that $\kappa$ increases (and $\sigma$ decreases) for stars with stronger differential rotation.}
\label{kappa_diffrot}
\end{figure}

In \Fref{kappa_diffrot} we show the variation of $\kappa$ as a function of the compactness $M/R$ for sequences of stars with different values of $\hat{A}^{-1}$ and a representative fixed value of the total angular momentum. We can see that the effect of differential rotation acts to increase $\kappa$, and it is more pronounced for more compact stars, leading to a correction that ranges from 3\% to 22\% (this trend reflects the behavior shown in \Fref{6-models}). 

\begin{figure}[!htb]
\begin{center}
\includegraphics[width=0.98\columnwidth]{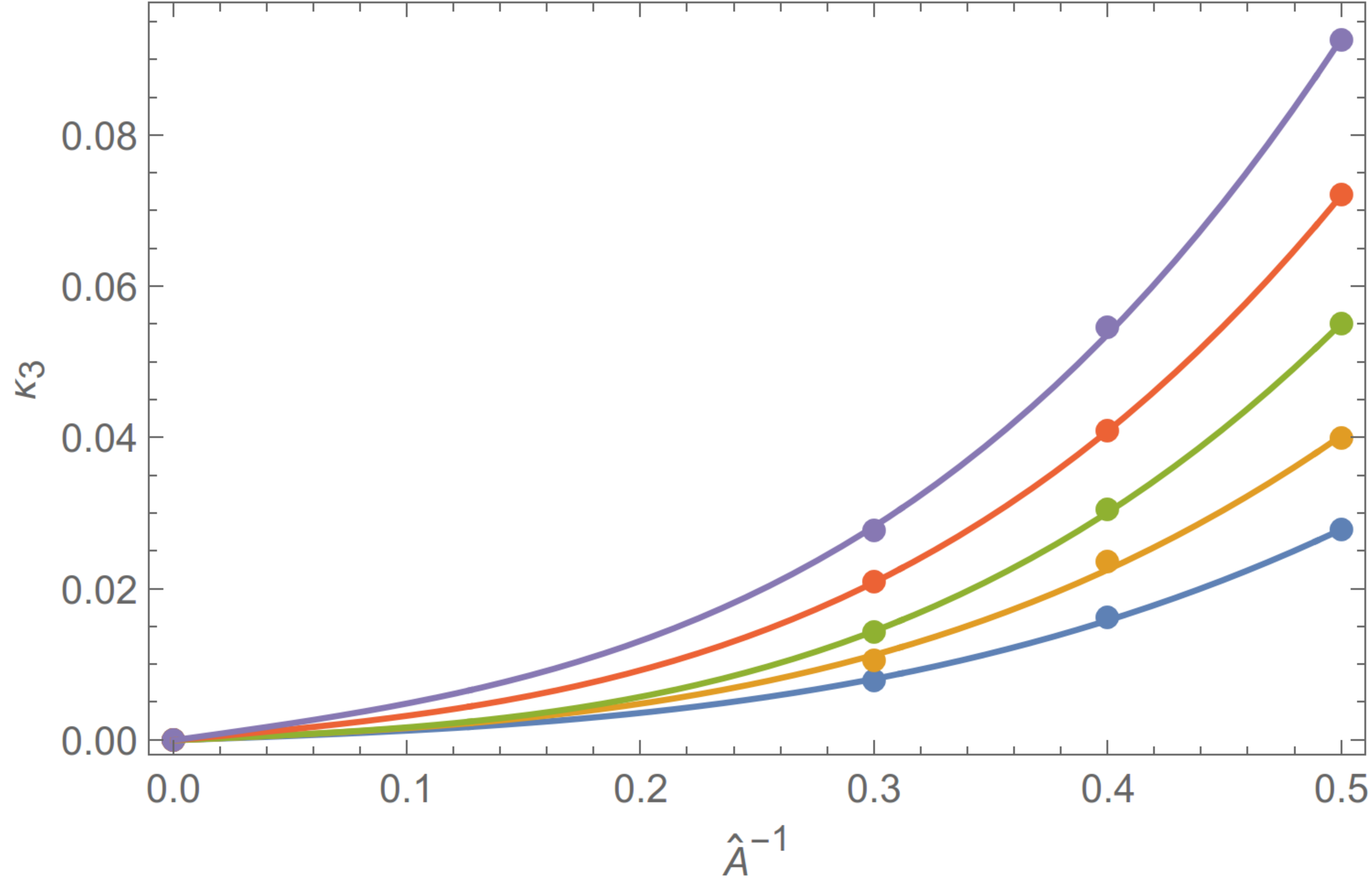}
\end{center}
\caption{$\kappa_3 \times \hat{A}^{-1}$ for sequences of stars with different values of $M/R$, which were obtained by interpolation from the data in \Fref{kappa_diffrot}. The solid lines represent fits of the form $\kappa_3 = c\hat{A}^{-1}+ d\hat{A}^{-3}$. From top to bottom, $M/R = 0.2, 0.18,...\,,0.12$. The differential rotation correction $\kappa_3$ is more significant for more compact stars.}
\label{kappa3}
\end{figure}

We define here the differential rotation correction $\kappa_3$ to the value of $\kappa$ in the simplest way possible by a generalization of \Eref{eq:kappaxrot}:
\begin{equation}
  \kappa = \kappa_0 + \kappa_2\, \frac{\Omega^2}{\pi G \bar{\rho}_0} + \kappa_3\,,
  \label{eq:kappaxdiffrot}
\end{equation}
and $\kappa_3$ is shown as a function of $\hat{A}^{-1}$ in \Fref{kappa3} for the same data that was presented in \Fref{kappa_diffrot}. We found that $\kappa_3$ is best described by a 2-parameter fit of the form 
\begin{equation}
 \kappa_3 = c\hat{A}^{-1}+ d\hat{A}^{-3}\,,
\label{eq:k3}
\end{equation}
where the parameters $c$ and $d$ will depend in general on the compactness and total angular momentum of the star (but more strongly on the compactness, see the discussion below and the Appendix). For example, we find $c = 0.0425$ and $d = 0.57$ for the model with $M/R = 0.2$ in \Fref{kappa3}.

\begin{figure}[!htb]
\begin{center}
\includegraphics[width=0.98\columnwidth]{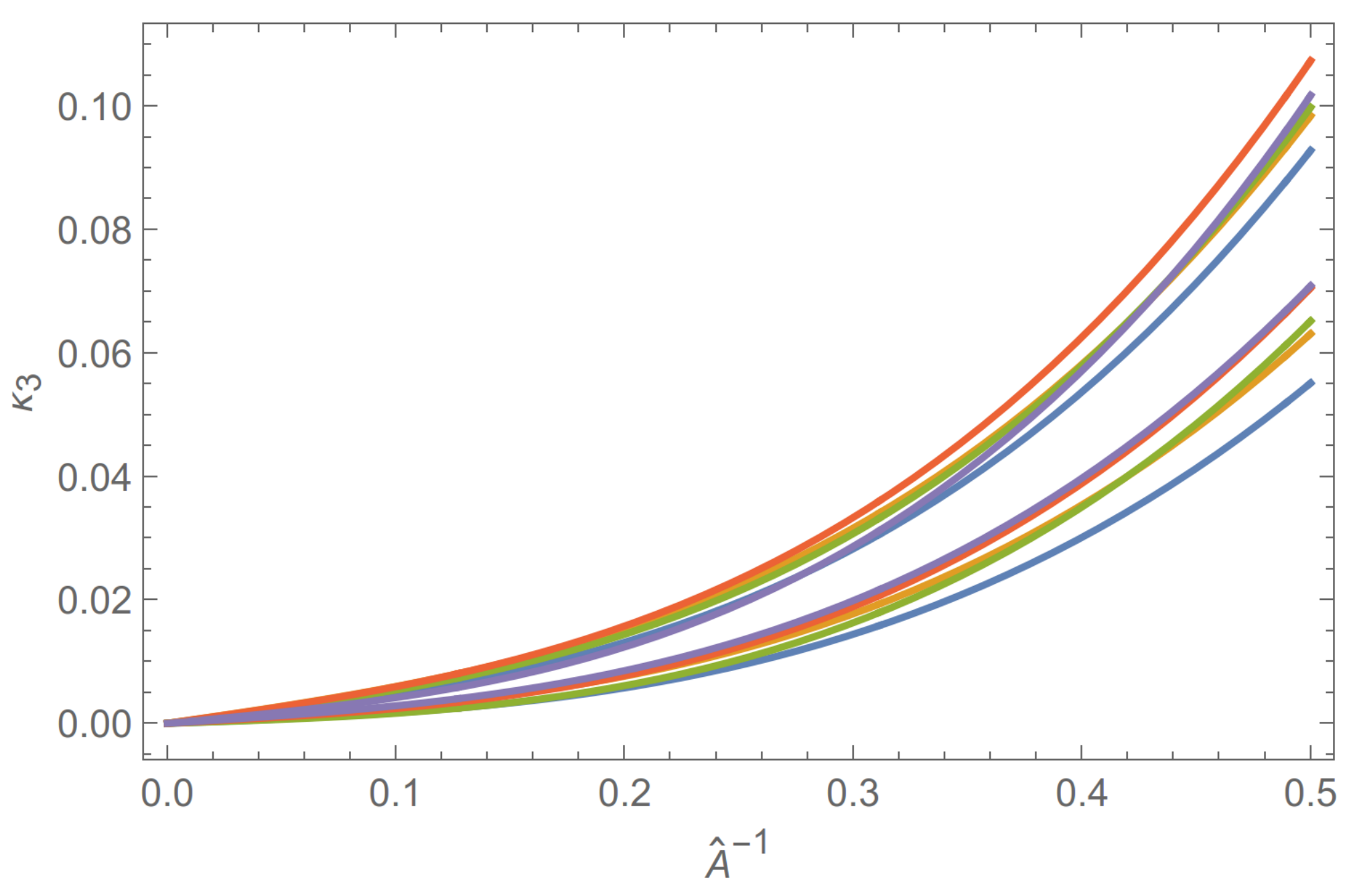}
\end{center}
\caption{$\kappa_3 \times \hat{A}^{-1}$ for sequences of stars with different values of $f$ ranging from $200-600\,\textrm{Hz}$ and two different values of compatness: $M/R = 0.20$ (upper curves) and $M/R = 0.14$ (lower curves). In this range, $\kappa_3$ is almost independent of the total angular momentum of the star, and depends much more strongly on the compactness. (The color code is \{blue, yellow, green, red, violett\} for $f=600,500,400,300,200\, \textrm{Hz}$ respectively.)}
\label{kappa3-all}
\end{figure}

We have also investigated how $\kappa_3$ depends on the total angular momentum of the star, and the results are shown in \Fref{kappa3-all}. The different curves represent sequences of stars that have different values of total angular momentum, and they naturally split according to their compactness. The upper curves have a fixed high compactness, while the lower curves have a fixed low compactness. Comparing the two sets of data we can see that $\kappa_3$ depends much more strongly on the compactness than it does on the total angular momentum of the star.

\section{Conclusion}
\label{sec:conclusion}

Here we have presented a detailed study on the influence of both fast rotation and differential rotation on the r-mode frequency of compact stars, analyzing the deviation from the well-known Newtonian value $\kappa = 2/3$.

We have studied for the first time the dependence of the rotational correction $\kappa_2$ on the compactness $M/R$ of the stars. We found that $\kappa_2$ is very sensitive to the compactness of the star. It decreases with increasing compactness, like the slow rotation term $\kappa_0$, and the difference between values obtained for models with small and large compactness is $\sim 70\%$. 
 
Comparing our fit for $\kappa_0$ with the result from the full general relativistic analysis of \cite{Idrisy} for low rotation rates, we were able to show a quantitative assessment of the accuracy of the Cowling approximation for r-modes. Our values, obtained within the Cowling approximation, are shifted to lower frequencies by $6\%-11\%$, in good agreement with the expectation of good accuracy for Cowling results for r-mode frequencies.  

For pulsars in LMXB rotational effects on the r-mode frequency can be strong
and larger than relativistic effects, in particular, for rapidly spinning stars
with low compactness. For a possible guided band-limited gravitational wave 
search of a known LMXB this would imply that the band is broadened at the lower end.

Differential rotation may be present in LMXB pulsars and it contributes to further increase $\kappa$. The differential rotation correction $\kappa_3$ is more sensitive to the compactness than to the rotation of the star and it increases with the compactness (contrary to $\kappa_0$ and $\kappa_2$) for sequences of stars with constant angular momentum, exceeding 20\% for high compactness models. This suggests that even a small degree of differential rotation should not be neglected for stars with very high compactness.

We plan to continue this work to study the effects of fast rotation and differential rotation on the gravitational wave \cite{Ipser} and viscosity damping timescales \cite{Lindblom2} and the r-mode instability window.

\acknowledgments
We thank Wolfgang Kastaun and Cole Miller for useful discussions and comments. This work was supported by the S\~ao Paulo Research Foundation (FAPESP grant 2015/20433-4) and by the Max Planck Society.

\appendix*
\section{The differential rotation correction $\kappa_3$ as a function of the compactness and total angular momentum of the star}
\label{sec:appendix}

\begin{figure}[!htb]
\begin{center}
\includegraphics[width=0.98\columnwidth]{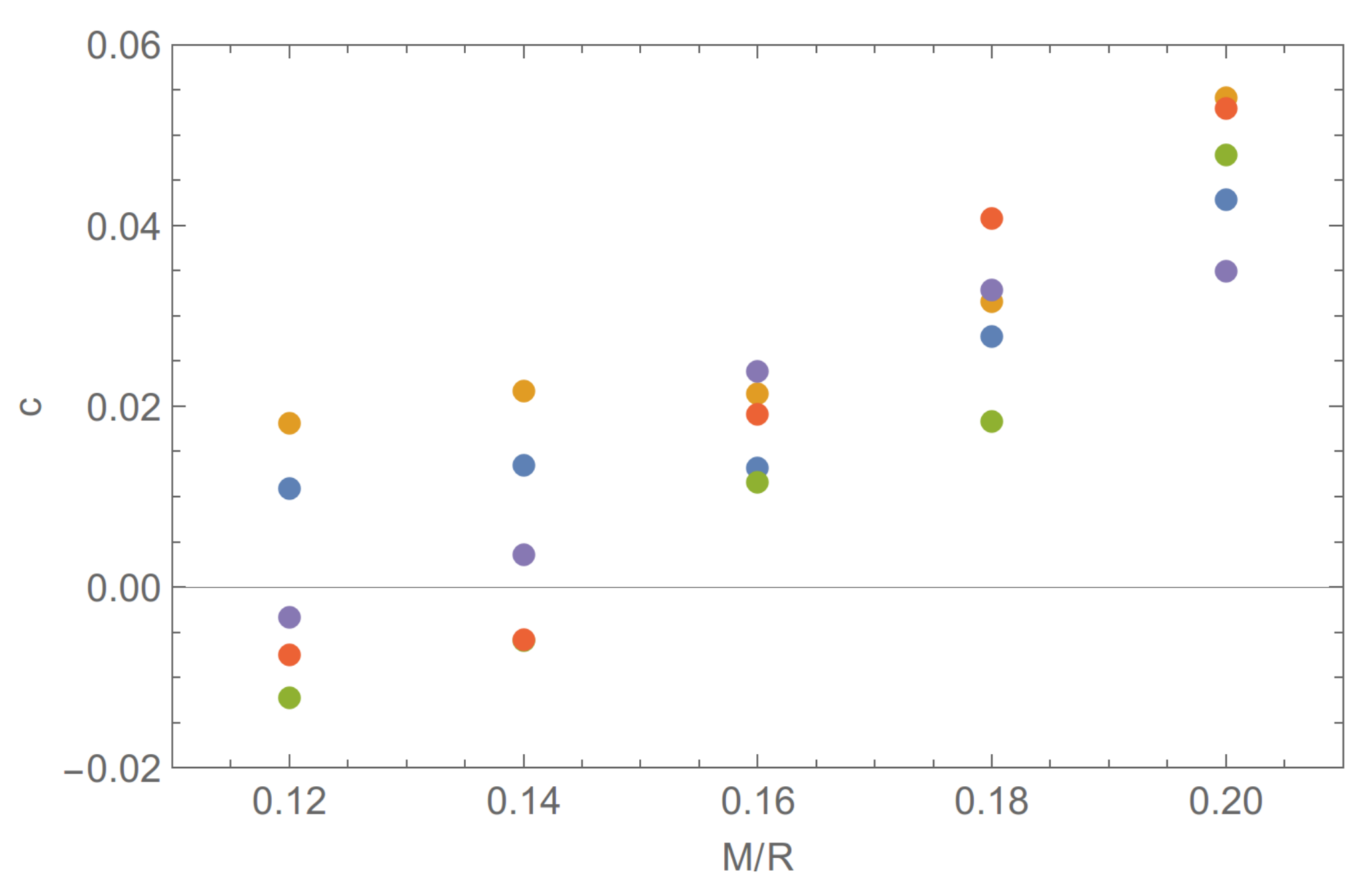}
\end{center}
\caption{$c \times M/R$ as defined by \Eref{eq:k3} for sequences of stars with different values of $f$ ranging from $200-600\,\textrm{Hz}$. The color code is the same as in \Fref{kappa3-all}.}
\label{c}
\end{figure}

\begin{figure}[!htb]
\begin{center}
\includegraphics[width=0.98\columnwidth]{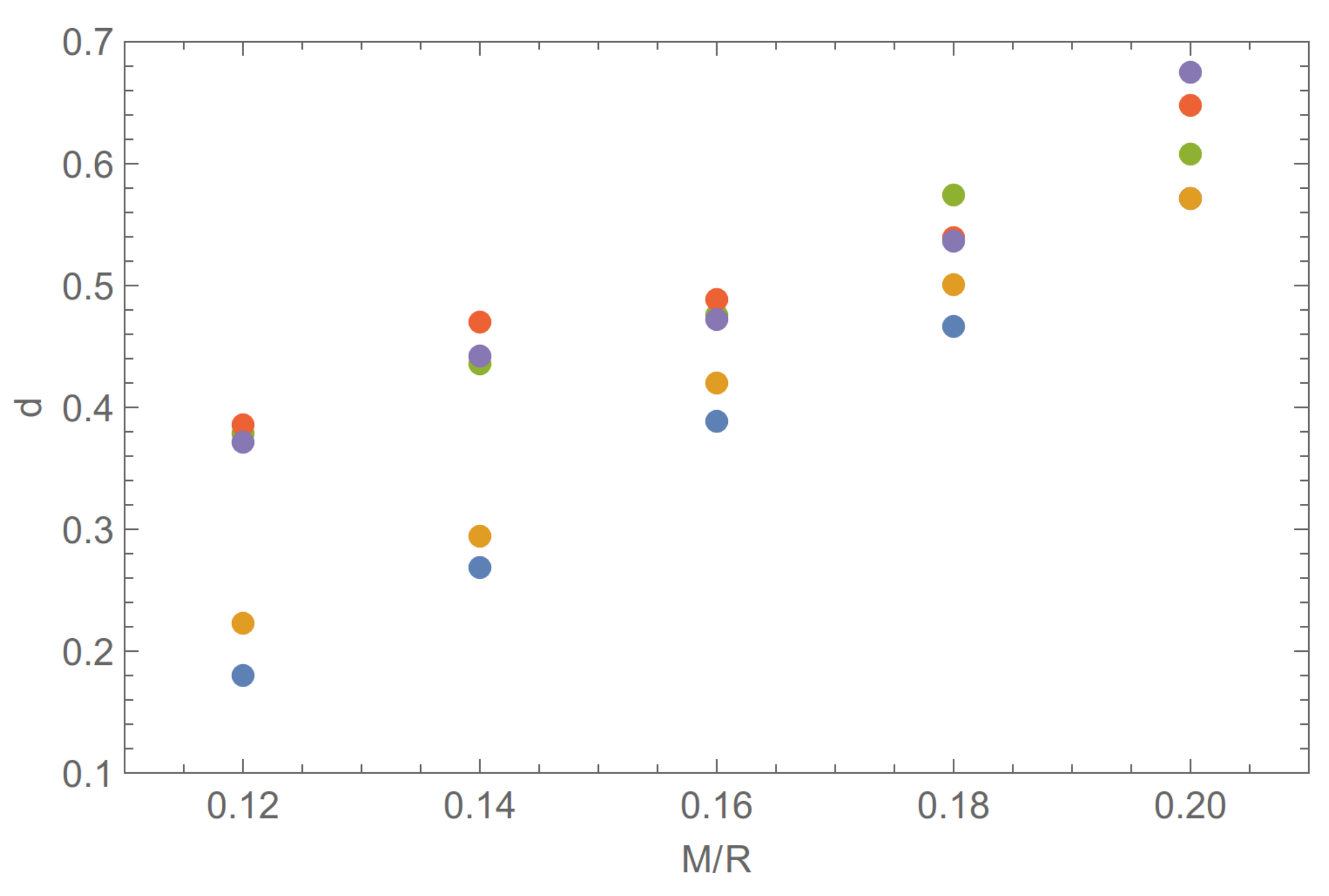}
\end{center}
\caption{$d \times M/R$  as defined by \Eref{eq:k3} for sequences of stars with different values of $f$ ranging from $200-600\,\textrm{Hz}$. The color code is the same as in \Fref{kappa3-all}.}
\label{d}
\end{figure}

We present in \Fref{c} and \Fref{d} the values obtained for the coefficients $c$ and $d$ defined by \Eref{eq:k3}
 to parametrize the differential rotation correction $\kappa_3$, defined in \Eref{eq:kappaxdiffrot}. 

As we can see from both figures, the coefficients depend more strongly (and approximately linearly) on the compactness, with some spread caused by the variation of the angular momentum (and numerical errors), which becomes more relevant for lower compactness $M/R \lesssim 0.14$, as was already expected from \Fref{kappa3-all}. We also expect that numerical errors will increase for more slowly rotating stars, which would need longer runs for a more accurate determination of their (lower) r-mode frequency.

For higher compactness models with $M/R \gtrsim 0.14$ the spread caused by the variation in the total angular momentum we considered is $(c_{\rm min}-c_{\rm max})/c_{\rm min}\approx 60\%$ and $(d_{\rm min}-d_{\rm max})/d_{\rm min}\approx 20\%$.

\begin{thebibliography}{99}

  \bibitem{Chandrasekhar} S. Chandrasekhar, Phys. Rev. Lett. {\bf 24}, 611 (1970).
  
  \bibitem{Friedman} J. L. Friedman and B. F. Schutz, Astrophys. J. {\bf 222}, 281 (1978).
  
  \bibitem{Papaloizou} J. Papaloizou, J. E. Pringle, Mon. Not. R. Astron. Soc. {\bf 182}, 423 (1978).

  \bibitem{Owen} B. J. Owen, Phys. Rev. D {\bf 82}, 104002 (2010).
  
  \bibitem{Arras} P. Arras {\it et al.}, Astrophys. J. {\bf 591}, 1129 (2003). 
  
  \bibitem{Bondarescu1} R. Bondarescu, S. A. Teukolsky and I. Wasserman, Phys. Rev. D {\bf 76}, 064019 (2007).
  
  \bibitem{Bondarescu2} R. Bondarescu, S. A. Teukolsky and I. Wasserman, Phys. Rev. D {\bf 79}, 104003 (2009).
  
  \bibitem{Haskell} B. Haskell, K. Glampedakis and N. Andersson, Mon. Not. R. Astron. Soc. {\bf 441}, 1662 (2014).
  
  \bibitem{Strohmayer1} T. Strohmayer and S. Mahmoodifar, Astrophys. J. Lett. {\bf 793}, L38 (2014).

  \bibitem{Strohmayer2} T. Strohmayer and S. Mahmoodifar, Astrophys. J. {\bf 784}, 72 (2014).

  \bibitem{Manchester} R. N. Manchester, G. B. Hobbs, A. Teoh and M. Hobbs, Astron. J. {\bf 129}, 1993 (2005).

  \bibitem{Rezzolla1} L. Rezzolla, F. K. Lamb, D. Markovic and S. L. Shapiro, Phys. Rev. D {\bf 64},104013 (2001).
  
  \bibitem{Rezzolla2} L. Rezzolla, F. K. Lamb, D. Markovic and S. L. Shapiro, Phys. Rev. D {\bf 64},104014 (2001).
  
  \bibitem{Kastaun1} W. Kastaun, R. Ciolfi and B. Giacomazzo, Phys. Rev. D {\bf 94}, 044060 (2016).

  \bibitem{Stavridis} A. Stavridis, A. Passamonti and K. Kokkotas, Phys. Rev. D {\bf 75}, 064019 (2007).

  \bibitem{Passamonti1} A. Passamonti, A. Stavridis and K. Kokkotas, Phys. Rev. D {\bf 77}, 024029 (2008).

  \bibitem{Chirenti} C. Chirenti, J. Skakala and S. Yoshida, Phys. Rev. D {\bf 87}, 044043 (2013).
  
  \bibitem{Krueger} C. Kr\"uger, E. Gaertig and K. Kokkotas, Phys. Rev. D {\bf 81}, 084019 (2010).

  \bibitem{Lockitch} K. H. Lockitch, J. L. Friedman and N. Andersson, Phys.Rev. D {\bf 68}, 124010 (2003).

  \bibitem{Idrisy} A. Idrisy, B. J. Owen and D. I. Jones, Phys. Rev. D {\bf 91}, 024001 (2015).
  
  \bibitem{Ferrari} V. Ferrari, L. Gualtieri and S. Marassi, Phys. Rev. D {\bf 76}, 104033 (2007).

  \bibitem{Stergioulas} N. Stergioulas, Living Rev. Relativity {\bf 6}, 3 (2003).
  
  \bibitem{Lindblom1} L. Lindblom, G. Mendell, B. J. Owen, Phys. Rev. D {\bf 60}, 064006 (1999).

  \bibitem{Passamonti2} A. Passamonti, B. Haskell and N. Andersson, Mon. Not. R. Astron. Soc. {\bf 396}, 951 (2009).

    \bibitem{Komatsu1} H. Komatsu, Y. Eriguchi and I. Hachisu, Mon. Not. R. Astron. Soc. {\bf 237}, 355 (1989).
  
  \bibitem{Komatsu2} H. Komatsu, Y. Eriguchi and I. Hachisu, Mon. Not. R. Astron. Soc. {\bf 239}, 153 (1989).

  \bibitem{Galeazzi} F. Galeazzi, S. Yoshida and Y. Eriguchi, Astron. \& Astrophys. {\bf 541}, A156 (2012).

  \bibitem{Mach} P. Mach and E. Malec, Phys. Rev. D {\bf 91}, 124053 (2015).

  \bibitem{Baumgarte} T. W. Baumgarte, S. L. Shapiro and M. Shibata, Astrophys. J {\bf 528}, L29 (2000).

  \bibitem{ET} \url{http://www.einsteintoolkit.org}

  \bibitem{Pollney} Denis Pollney et al., Phys.Rev. D {\bf 83}, 044045 (2011).

  \bibitem{Jasiulek} M. Jasiulek, Class. Quantum. Grav. {\bf 29}, 015008 (2012).
  
  \bibitem{Lorene} \url{http://www.lorene.obspm.fr}

  \bibitem{Ruoff} J. Ruoff, A. Stavridis and K. Kokkotas, Mon. Not. R. Astron. Soc. {\bf 339}, 1170 (2003).

  \bibitem{Kastaun2} W. Kastaun, Phys. Rev. D {\bf 84}, 124036 (2011).

  \bibitem{Yoshida} S. Yoshida and Y. Kojima, Mon. Not. R. Astron. Soc. {\bf 289}, 117 (1997).

  \bibitem{Ipser} J. Ipser and L. Lindbom, Astrophys. J. {\bf 373}, 213 (1991).

  \bibitem{Lindblom2} L. Lindblom, B. Owen and S. Morsink, Phys. Rev. Lett {\bf 80}, 4843 (1998).

\end{thebibliography}
\end{document}